\UseRawInputEncoding
\documentclass[aip,apl,reprint]{revtex4-1}

\usepackage[pdftex]{graphicx}
\graphicspath{{Figures/}}

\renewcommand{\deg}{^\circ}

\begin{document}

\title{Active Sites of Te-hyperdoped Silicon by Hard X-ray Photoelectron Spectroscopy}

\author{Moritz Hoesch}
\affiliation{Deutsches Elektronen-Synchrotron DESY, Notkestr. 85, 22607 Hamburg, Germany}

\author{Olena Fedchenko}
\affiliation{Johannes Gutenberg-Universit\"at, Institut f\"ur Physik, 55128 Mainz, Germany}

\author{Mao Wang}
\affiliation{Helmholtz-Zentrum Dresden-Rossendorf, Institute of Ion Beam Physics and Materials Research, Bautzner Landstraße 400, 01328 Dresden, Germany}

\author{Christoph Schlueter}
\affiliation{Deutsches Elektronen-Synchrotron DESY, Notkestr. 85, 22607 Hamburg, Germany}

\author{Dmitrii Potorochin}
\affiliation{Deutsches Elektronen-Synchrotron DESY, Notkestr. 85, 22607 Hamburg, Germany}
\affiliation{TU Bergakademie Freiberg, Institute of Experimental Physics, 09599 Freiberg, Germany}

\author{Katerina Medjanik}
\affiliation{Johannes Gutenberg-Universit\"at, Institut f\"ur Physik, 55128 Mainz, Germany}

\author{Sergey Babenkov}
\affiliation{Johannes Gutenberg-Universit\"at, Institut f\"ur Physik, 55128 Mainz, Germany}

\author{Anca S. Ciobanu}
\affiliation{Deutsches Elektronen-Synchrotron DESY, Notkestr. 85, 22607 Hamburg, Germany}

\author{Aimo Winkelmann}
\affiliation{Academic Centre for Materials and Nanotechnology (ACMiN), AGH University of Krakow, 30-059 Krak\'{o}w, Poland}

\author{Hans-Joachim Elmers}
\affiliation{Johannes Gutenberg-Universit\"at, Institut f\"ur Physik, 55128 Mainz, Germany}

\author{Shengqiang Zhou}
\affiliation{Helmholtz-Zentrum Dresden-Rossendorf, Institute of Ion Beam Physics and Materials Research, Bautzner Landstraße 400, 01328 Dresden, Germany}

\author{Manfred Helm}
\affiliation{Helmholtz-Zentrum Dresden-Rossendorf, Institute of Ion Beam Physics and Materials Research, Bautzner Landstraße 400, 01328 Dresden, Germany}

\author{Gerd Sch\"onhense}
\affiliation{Johannes Gutenberg-Universit\"at, Institut f\"ur Physik, 55128 Mainz, Germany}

\date{\today}

\begin{abstract}

Multiple dopant configurations of Te impurities in close vicinity in silicon are investigated using photoelectron spectroscopy, photoelectron diffraction, and Bloch wave calculations. The samples are prepared by ion implantation followed by pulsed laser annealing. The dopant concentration is variable and high above the solubility limit of Te in silicon. The configurations in question are distinguished from isolated Te impurities by a strong chemical core level shift. While Te clusters are found to form only in very small concentrations, multi-Te configurations of type dimer or up to four Te ions surrounding a vacancy are clearly identified. For these configurations a substitutional site location of Te is found to match the data best in all cases. For isolated Te ions this matches the expectations. For multi-Te configurations the results contribute to understanding the exceptional activation of free charge carriers in hyperdoping of chalcogens in silicon.

\end{abstract}

\maketitle

corresponding author: Moritz Hoesch $<$moritz.hoesch@desy.de$>$


Doping of the band semiconductor silicon follows well-established principles: to enhance hole (electron) concentration, elements of group 13 (group 15) of the periodic table are introduced leading to singly ionised impurities.~\cite{pearson49} Doping to very high limits can lead to a transition to a metallic state above the insulator-to-metal transition (IMT). For these singly ionised dopants the IMT matches well with the expectations from the band model of Mott.~\cite{Mott74} For doubly ionised dopants of group 16, one column further in the periodic table, a similar transition criterion holds, as studied in the case of Si:Se\cite{ertekin2012PRL} and  Si:Te.\cite{wang2020}
Above the IMT, the free carrier concentration is in principle freely tuneable by the dopant concentration. However, the mutual interaction of dopant atoms may modify their wave functions and their ability to contribute free carriers, which depends also on the local structure surrounding the dopant as well as structural distortions that may occur as two or more of the dopant atoms come in close vicinity.~\cite{voyles2002atomic,Rummukainen2005PRL} For almost all singly ionising \emph{n}-type doped silicon studied to date, the formation of donor pairs,  clusters or  defect complexes results in an intrinsic upper limit at around 5$\times$10$^{20}$ cm$^{-3}$ for the electron concentration.~\cite{Rummukainen2005PRL,Chadi1997PRL,Ranki2002PRL,Pinacho2005APL} 

Surprisingly, hyperdoping of silicon by Te results in electron concentration up to 8.3$\times$10$^{20}$ cm$^{-3}$ without saturation.~\cite{wang2019,wang2020} It can be assumed that the double-ionised nature of Te in silicon contributes to continued activation also of multi-Te complexes. A part of the charge may be trapped, but in contrast to single-ionised donors part of the charge remains active even in dimers or vacancy configurations. This has been studied theoretically for Se in silicon.\cite{Debernardi2021,Debernardi2022} Chalcogen ions are deep level donors in silicon and have low thermal solubilities.~\cite{Sanchez2010PRB} Hyperdoping refers to doping above the solubility limit. It is realised by non-thermal equilibrium methods, {\em e.g.} combining ion implantation and pulsed laser melting at nanoseconds.~\cite{ertekin2012PRL,wang2019,warrender2016laser} Potential applications of the hyperdoped silicon include the use as extended photodetectors in the infrared wavelength range.~\cite{said2011extended,warrender2016laser,jin2019high,wang2021silicon,rosseel2019characterization}

In this letter we address the identification of Te sites within the Si matrix. These are observed in samples of Si:Te at high doping concentration. We combine the method of photoelectron spectroscopy, which is sensitive to the electronic configuration of the dopant atom and its local surrounding, with the analysis of diffraction patterns of these photoelectrons that reveal the geometric configuration of the dopant site. This virtue of the method has also been used in other impurity systems.\cite{medjanik2021, fedchenko2022}  For the same crystal class, at lower photoelectron kinetic energy, studies have been performed on  Si:As\cite{tsutsui2017} and diamond:P\cite{yokoya2019} and summarised in a recent review.\cite{yokoya2022} We find that the vast majority of Te atoms are located on substitutional sites, which continues to be the case when Te in isolated monomers become the minority against dimer or more complex configurations of nearby Te impurities. For these, model cases are identified thus rationalising the ability of complex Te-Si configurations to contribute actively to the conductivity of Si above the IMT.


Samples of Si:Te with  peak Te concentrations up to 4.5~at\% have been prepared by ion implantation followed by pulsed laser melting.~\cite{wang2019}  Te impurities are located in a layer of approx. 100~nm thickness near the surface~\cite{wang2018}. The experiments were performed using X-ray photoelectron spectroscopy at beamline P22 of the storage ring X-ray source PETRA III ~\cite{schlueter2019} at a high photon energy and thus photoelectron kinetic energy (around 5690 eV for most data). Due to an inelastic mean free path of approx. 5~nm in this regime,\cite{Seah79} the probing depth can be as high as 20 nm and reaches into a region of near peak Te concentration. The top surface of the Si(001) wafers was etched just prior to the experiments using diluted hydrofluoric acid (HF at 10\%) followed by a pure ethanol wash and rapid introduction into the vacuum system. The samples were held at the temperature of 40 K during data taking.

\begin{figure}[!]
\begin{centering}
\includegraphics[width=\columnwidth]{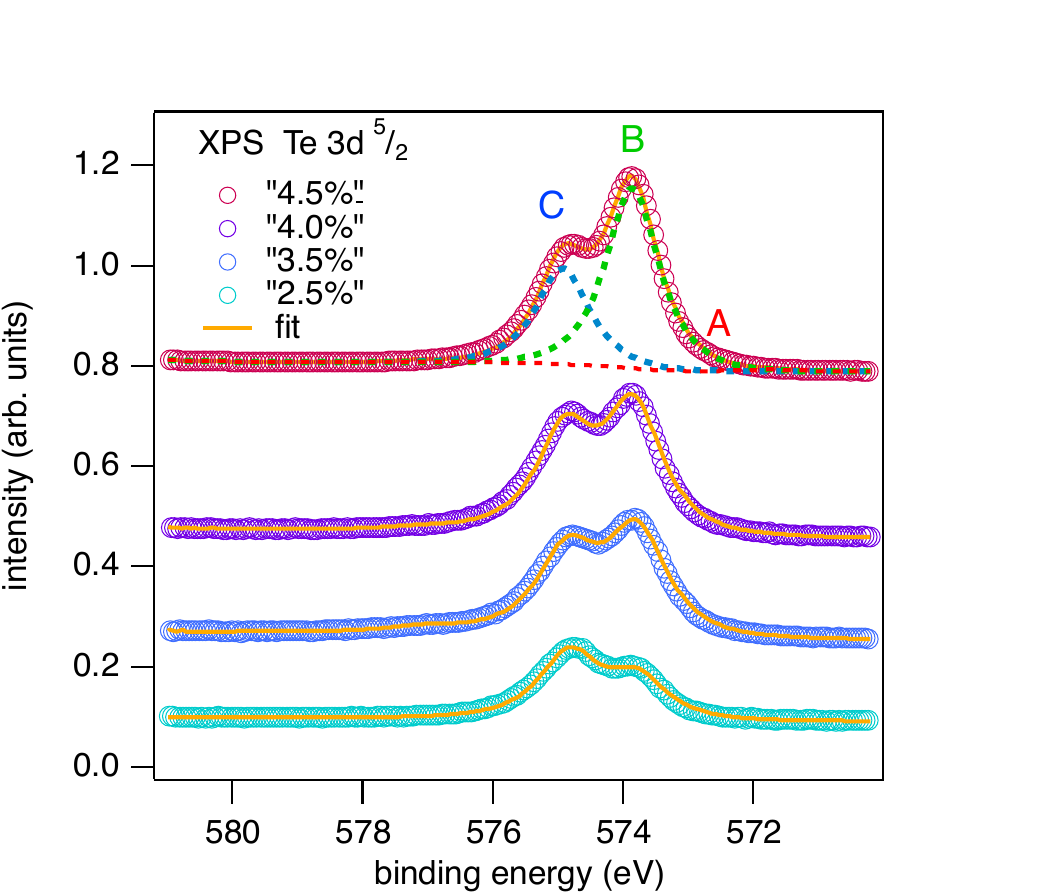}
\caption{X-ray photoelectron spectra of samples with varying Te nominal concentration in the Te 3d$_{5/2}$ region. 
The data for nominal concentration of 2.5, 3.5, and 4.0 \% are scaled as acquired with a hemispherical photoemission analyser and the increasing Te content is visible through increasing total peak area. The data for ''4.5~\%`` (acquired with the time-of-flight spectrometer) are scaled to match. Offsets are introduced for graphical clarity. The best fit spectra are shown as orange lines as described in the text. For ''4.5~\%`` the separate contributions from peak C and background, peak B, and peak A are shown as red, green, and blue dashed lines, respectively. }
\label{Fig:1}
\end{centering}
\end{figure}

Spectra  of  X-ray photoelectron spectroscopy (XPS) are shown in Fig.~\ref{Fig:1} for  samples of nominal Te concentration between 2.5 at\% and 4.5 at\%. The  peak heights related to Te is found to increase accordingly.
A multi-peak structure of at least two chemically shifted peaks is found in all samples in the region of  Te 3d$_{5/2}$. Chemical shifts in XPS can derive partially from the charge state of the probed ion and from its electrostatic environment. Further and often strong shifts are derived from the dynamical screening of the core hole created in photoemission due to quasi-free and metallic charge carriers. This screening usually results in an observed reduced  binding energy. The peak width of each chemically shifted component is ultimately limited by the uncertainty introduced by core hole life time and the final spectrum is composed of the sum of these overlapping contributions. A peak fitting analysis was performed using the CasaXPS software.\cite{fairley2021} Details of this analysis and numerical results are shown in supplemental material.  This multi-peak structure is indicative of a variation of  impurity ion configurations in the highly doped samples. The peak at lower binding energy (B), grows more rapidly than the peak at 1~eV higher binding energy (C). At the lowest doping concentration, where mostly substitutional site occupation of Te has previously been identified,~\cite{wang2019} peak B is significantly smaller than peak C. Intensity C is thus related to substitutional Te monomers, {\it i. e.} Te ions surrounded by at least one shell of Si in a lattice that is only locally distorted.

\begin{figure}[!]
\begin{centering}
\includegraphics[width=\columnwidth]{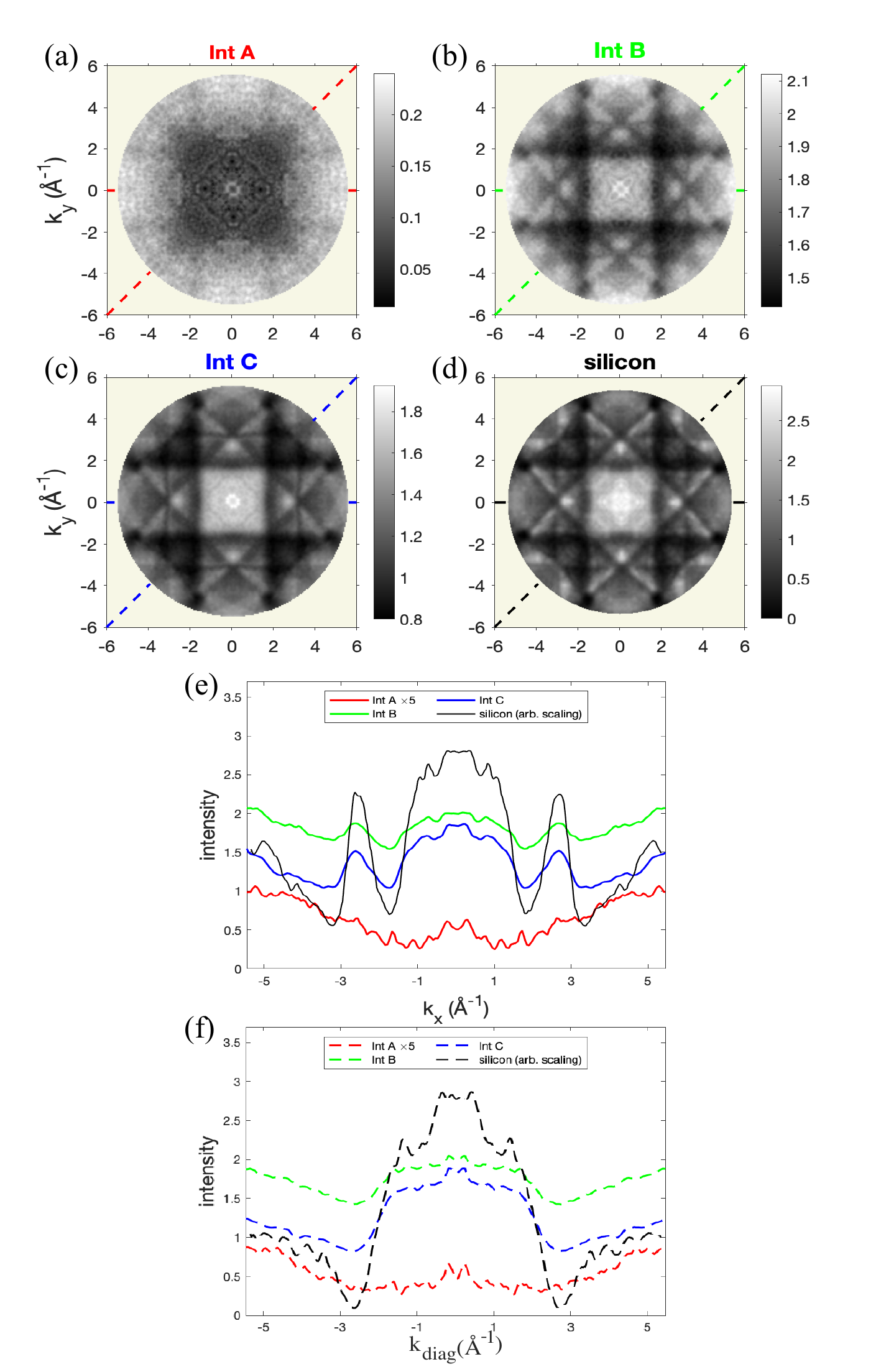}
\caption{(a-d) hXPD patterns recorded for the sample ''4.5~\%``. The patterns Int A - Int C correspond to a decomposition of binding energies as indicated in Fig.~\ref{Fig:1}. The pattern for silicon was recorded for Si 2p at the same kinetic energy as the Te 3d$_{5/2}$ level. The line plots in (e, f) show cuts through the centres of all patterns, as indicated by coloured dashes in (a-d).}
\label{Fig:2}
\end{centering}
\end{figure}

We now consider the high-energy photoelectron diffraction (hXPD) patterns generated by emission from these chemically shifted species. The data have been acquired using time-of-flight momentum microscopy~\cite{medjanik2018} for the highest doped sample ''4.5~\%``. For photoemission from a Si 2p core level a pattern of this kind is shown in Fig.~\ref{Fig:2}(d). Note that the photon energy was tuned to excite the electrons from the Si core level to the same $E_{kin}= 5690$~eV as from the Te 3d$_{5/2}$ level.  The  Kikuchi bands of enhanced photoemission intensity as well as characteristic low-intensity trenches are well-described by electron diffraction calculations assuming a point-like emitter.~\cite{fedchenko2020} Since each pixel of these patterns is related to a full spectrum, albeit at reduced counting statistics, we can resolve the  diffraction patterns for each chemical shift. These are shown in Fig.~\ref{Fig:2}(a-c) for Te 3d. Three distinct patterns are extracted by a pixel-by-pixel decomposition that extracts the intensities of three Gaussian peaks of equal widths and fixed positions.

All patterns follow the four-fold symmetry of the Si(001) surface. The patterns have thus been symmetrised after checking the four-fold structure. The incidence of the photon beam was along the $k_x$ direction in these plots, or equivalently from $-k_x$ or $\pm k_y$ since the four-fold symmetrisation reduces the weak effect of directionality of the photon beam. The incidence angle was $78\deg$ from the surface normal and linearly polarised radiation was used. Pattern A is nearly featureless but the characteristic trenches parallel to the $k_x$ and $k_y$ axes are also seen in this pattern, at identical positions to the other patterns. Patterns B and C have a striking resemblance to emission from Si. For pattern C, this is expected as this binding energy region  belongs to substitutional monomers. Pattern B is of highest intensity, but its overall contrast is further reduced, which will be discussed below. Note that the Kikuchi  pattern is sensitive to a displacement of the emitter position with respect to the long range registered lattice site~\cite{winkelmann2013, fedchenko2020} and the monomer is located on the same high symmetry position as a Si atom. 

Calculations for the analysis of these patterns proceed in two steps. For electronic structure calculations we construct supercells of $3\times 3 \times 3$ cubic Si unit cells and introduce one, two or more Te atoms as well as vacancies. By {\em ab-initio} density functional theory (DFT)  we determine the total-energy-minimised configuration for these configurations, thus leading to force free relaxed cells. The DFT model parameters within wien2k~\cite{wien2k} are given in the supplemental material. Vacancy-free configurations that we have considered are a substitutional monomer, a substitutional dimer of nearest neighbours (dim$_{\left<111\right>}$) and a configuration of two substitutional Te atoms interspaced by a Si atom dim$_{\left<100\right>}$. Configurations with a vacancy and up to four Te atoms in direct vicinity to the vacancy are labeled Te$_n$Vac with $n= 1 -4$.  In a second step the hXPD patterns are calculated for the Te emitter atoms in these configurations by Bloch wave diffraction calculations.~\cite{winkelmann2008, fedchenko2020}

\begin{figure}[!]
\begin{centering}
\includegraphics[width=\columnwidth]{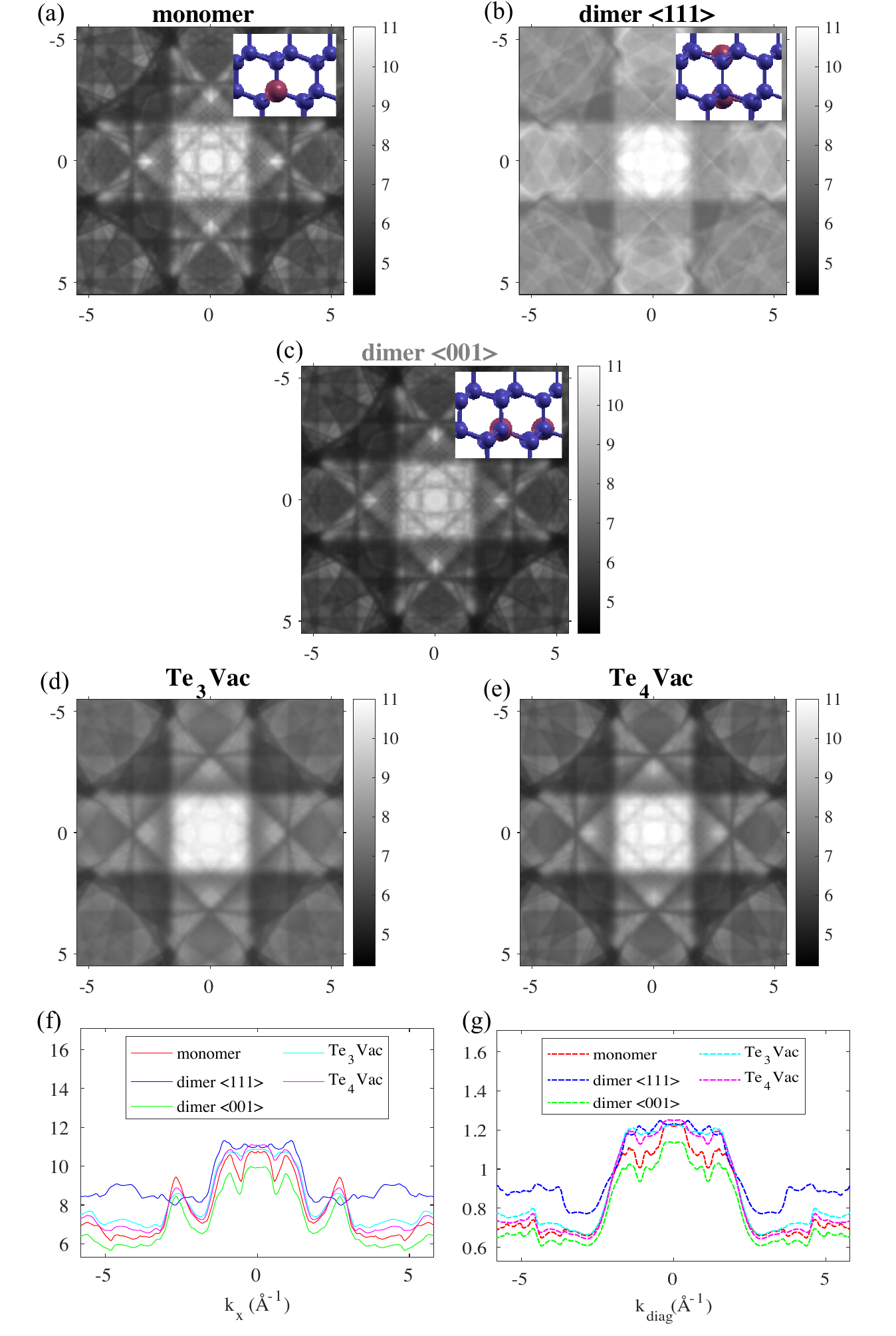}
\caption{(a-e) Calculated Bloch wave diffraction patterns for five local structure models as indicated above each panel. Inserts in (a-c) show the crystal structure viewed along $\left<110\right>$.
 The line plots shown in (f, g) are along the $k_y =0$ line and the $k_x = k_y$ line, respectively. }
\label{Fig:5}
\end{centering}
\end{figure}

The hXPD patterns for five selected configurations are shown in Fig.~\ref{Fig:5}.  On inspection we note the similarity of the calculations for the monomer and for the dim$_{\left<100\right>}$, Te$_3$Vac, and Te$_4$Vac configurations to each other. All patterns also compare well as to the experimental data of intensities (B) and (C).  The dim$_{\left<111\right>}$, as well as the configurations shown in supplemental material (Fig. S4)  match the data less well.


The diffraction patterns are generally well reproduced by calculations, with the exception of pattern A.  We assign this pattern to clusters of Te atoms with the low binding energy arising from Te atoms in clusters. Since pure Te is metallic, photoemission from these clusters will lead to a strong chemical shift to lower binding energies. The clusters may occur in various orientations with respect to the Si lattice and have a partially amorphous structure, thus leading to the near feature-less pattern. These Te atoms in clusters are a small minority of all Te atoms, given the very small spectral intensity.
 
A particularly good match is found for the monomer pattern C. Just the total contrast is smaller than that of the calculation, while  this contrast is well reproduced in calculations of pure Si.~\cite{fedchenko2020}  We assign this reduced contrast  to differences of vibrational amplitudes between  Si matrix atoms and a Te impurity that are not accounted for in our model. Also, any spill-over of intensity from the lower binding energy features B and A due to inelastic scattering will reduce the contrast in feature C. 

Feature B shows still lower contrast in the experimental data, while the best-matching calculation are dim$_{\left<100\right>}$, Te$_3$Vac, and  Te$_4$Vac.  
We have tested further configurations, namely Te$_1$Vac, and  Te$_2$Vac, but these also led to a rather poor match between calculation and pattern B.  A reasonable match is found for a tetragonal interstitial configuration, since this has the same local symmetry as a substitutional site (see also Ref.~[\onlinecite{fedchenko2020}]). This calculation is, however, not physical as it omits the DFT relaxation step. When relaxed, this structure assumes the hexagonal interstitial configuration, which has a different symmetry and a very poor match to any of the observed patterns.

We tentatively conclude that pattern B arises from a mixture of configurations dim$_{\left<001\right>}$, Te$_3$Vac, and  Te$_4$Vac that are present in the sample. Other configurations, namely dim$_{\left<111\right>}$, are  present at significantly smaller concentrations, at most a small fraction of all contributions to peak B.
They may exist as a minority and their contribution leads to a further reduction of the contrast of pattern B. The configurations Te$_3$Vac, and  Te$_4$Vac contain an isolated vacancy, evidence for which has also been found by recent positron annihilation experiments in samples of similar preparation.\cite{Shaikh2021} Theoretical support for a significant formation of multi-chalcogen vacancy configurations was found also in the case of Si:Se by first principles simulations, suggesting the likely formation of Te$_3$Vac and Te$_4$Vac.\cite{Debernardi2021}

For a quantitative analysis, we assert that peak C (34.5~\% of total XPS intensity) derives purely from monomers, and peak B (64.8~\% of XPS intensity) derives from multi-Te configurations.  Since each Te atom contributes equally to the photoemission signal, this allows to estimate that about 1/3 of Te in sample ''4.5 \%`` is in monomers and 2/3 of Te in multi-Te configurations (dim$_{\left<111\right>}$, dim$_{\left<001\right>}$, Te$_2$Vac,  Te$_3$Vac, and  Te$_4$Vac ). Te in clusters (peak A) is nearly negligible. Further assuming that each multi-Te impurity contains $2.5\pm0.4$ Te ions, the ratio of monomer impurities to multi-Te impurities is $1.38 \pm 0.22$.

\begin{figure}[!]
\begin{centering}
\includegraphics[width = 0.8 \columnwidth]{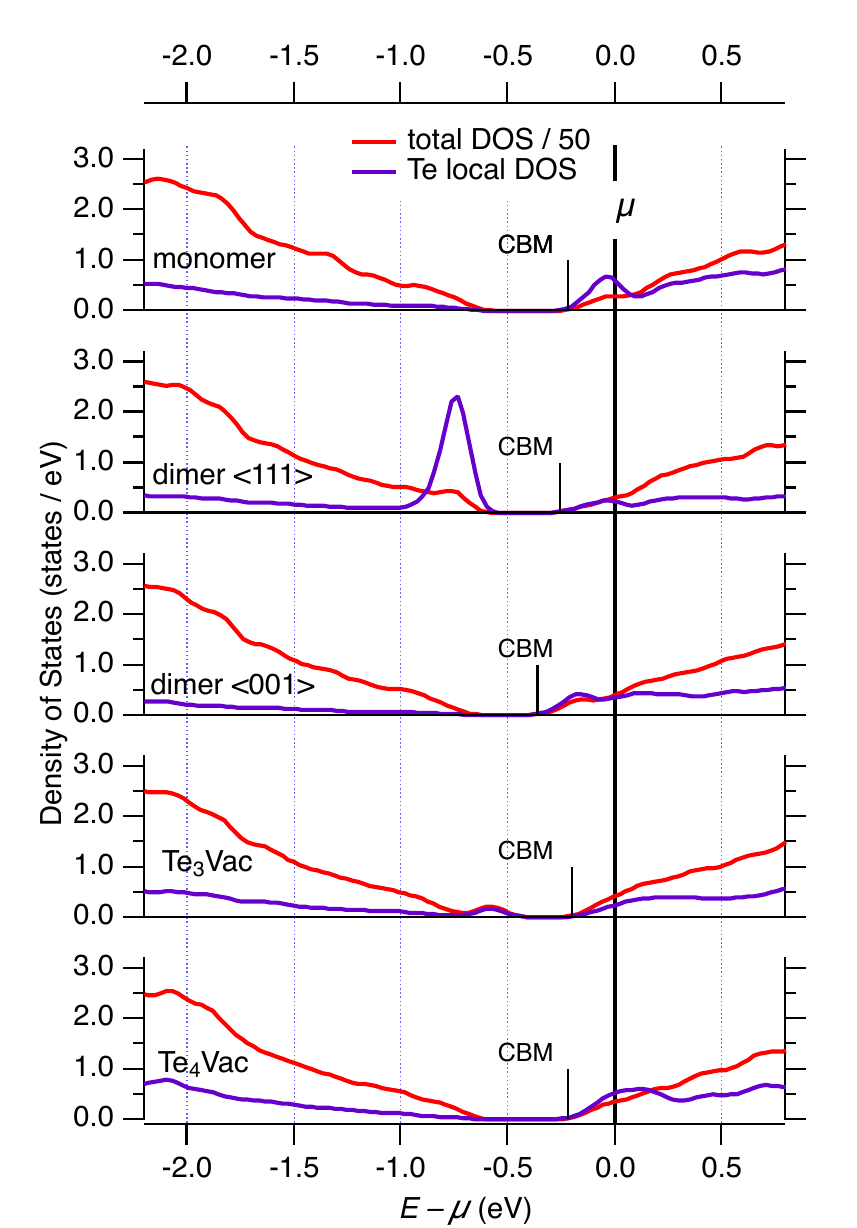}
\caption{Density of states and local density of states at the Te site, obtained from {\em ab-initio} calculated supercells with local structures as indicated. The energy is scaled relative to the chemical potential $\mu$. The conduction band minimum (CMB) is indicated for each case.}
\label{Fig:6}
\end{centering}
\end{figure}

The variation in binding energy can be due to a different charge state of the Te ions, or due to a difference in screening, or a combination of both. For pattern A we have already identified poorly ionised Te atoms in clusters leading to an electrostatic binding energy reduction. The configurations monomer, dim$_{\left<111\right>}$, dim$_{\left<001\right>}$, Te$_3$Vac, and  Te$_4$Vac on the other hand all correspond to strongly ionised Te. This is evident from the density of states (DOS) extracted from the calculations and shown in Fig.~\ref{Fig:6}. All configurations show a similar Fermi energy ($E-\mu$ of the conduction band minimum CBM) of around 0.3~eV. In addition, a shallow core level of Te at $E_B = 5.1$~eV is found at near identical energy in these calculations (shown in supplemental material Figure S6). The electrostatic environment of Te in these five configurations is thus similar and the difference in binding energy must derive from dynamics in the photoemission process. In dim$_{\left<111\right>}$, dim$_{\left<001\right>}$, Te$_3$Vac, and  Te$_4$Vac  the close vicinity of doubly ionised Te atoms and the increased local density of free carriers lead to increased screening when compared to the monomer case, which results in a reduced binding energy.

In summary we have observed a multi-peak structure of the Te core level Te 3d$_{5/2}$. A small component (A) of the spectra is identified as arising from Te clusters. The high binding energy component (C) is readily identified as belonging to substitutional monomers. Component (B) also gives a clear and similar photoelectron diffraction pattern. Calculations of selected multi-Te configurations of Te in the silicon matrix are compared to these experimental data. 
The resulting patterns for  dim$_{\left<111\right>}$, Te$_1$Vac and Te$_2$Vac do not match the data well. These multi-Te configurations may still be present in the sample, but as a minority. Dim$_{\left<100\right>}$ as well as Te$_3$Vac and Te$_4$Vac with vacancies give a good match to the experimental data. Therefore the data strongly support a large fraction of Te in such configurations and only a minority in regular dimers, or other configurations. 

Inspection of the model calculations in regards to the density of states shows that ionisation of Te is similar in all configurations, thus demonstrating the double-donor quality of Te also in the complex configurations. The high local density of free charge carriers is also held responsible for the chemical shift of the core level to lower binding energy. This strong ability of Te to supply electrons in both, monomer and multi-Te configurations serves to explain the previously observed saturation-free increase of metallic carrier density in Te-hyperdoped silicon.

\section*{Acknowledgements}

We wish to thank Fumihiko Matsui, Wolfgang Drube, Sanjoy Mahatha, Alberto Debernardi, and Mohd Saif Shaikh for discussions. We thank Dmytro Kutnyakhov for careful reading of our manuscript. Photoemission data from Beamline P22-HAXPES at PETRA-III of DESY contributed to the results presented here. We gratefully acknowledge financial support by BMBF (projects 05K19UM1, 05K19UM2, and 05K22OF2 within ErUM-Pro), by Deutsche Forschungsgemeinschaft (Transregio SFB 173 Spin+X 268565370, project A02, and individual Research Grants 445049905), and by the Polish National Science Centre (NCN) grant 2020/37/B/ST5/03669. The ion implantation was done at the Ion Beam Center at Helmholt-Zentrum Dresden-Rossendorf.

\section*{Author Declarations}

\subsection*{Conflict of Interest}

The authors have no conflicts to disclose.

\subsection*{Author Contributions}

MW, SZ, and MaHe made and characterised the sam- ples. ASC, DP, and MoHo prepared the surfaces for measurements. The measurements were conducted by CS, OF, SB, KM, HJE, and GS. Modelling calculations were performed by AM and OF. The study was conceived and designed by MoHo and GS. All authors contributed to the discussion of the results and review of the manuscript.

\section*{Data Availability}
The data that support the findings of this study are available from the corresponding authors upon reasonable request.

\section*{Supplemental Material}

Supplemental is provided in pdf format. The document contains XPS spectra over a larger binding energy range (Fig.~S1), as well as from additional sample preparations (Fig.~S2) and it details their fitting procedure and summarises the numerical results (Table~S1). An alternative, simpler data reduction method is introduced and the corresponding results shown (Fig.~S3). Modelling calculation results from cases incompatible with the experimental data are presented (Fig.~S4) and the DOS for these cases is shown along with the DOS for pure silicon (Fig.~S5). A further figure summarises the findings of DOS for all model cases over a high binding energy range (Fig.~S6). A table figure gives all numerical results of the XPS fitting procedure.

\section*{References}

\bibliography{Te-Si001}

\end{document}